\documentclass{ws-ijmpa}

\begin{document}

\markboth{J.L. Nagle}
{Heavy Ion Physics}

%
\catchline{}{}{}{}{}
%

\title{Heavy Ion Physics and Quark-Gluon Plasma}

\author{\footnotesize J.L. Nagle}

\address{University of Colorado at Boulder\\
Boulder, CO 80309, USA
}

\maketitle

\pub{Received (15 11 2004)}{Revised (15 11 2004)}

\begin{abstract}
These proceedings represent a brief overview of the exciting physics coming
out from the Relativistic Heavy Ion Collider (RHIC) at Brookhaven National Laboratory.
The experimental results from BRAHMS, PHOBOS, PHENIX and STAR indicate a strongly-coupled
state of matter that can only be described on the partonic level. We review 
some of the latest experimental results as we presented at the meeting of the Division of Particles and Fields
of the American Physical Society in Riverside, CA in August 2004.
\end{abstract}

\section{Introduction}	
The presentation began with the ``bottom line'' conclusions.  The Relativistic Heavy Ion Collider (RHIC) has been operational
since 2000 to study matter at extreme temperatures.  We can ask and answer a set of straightforward questions
about the program.  

(1)  Have the accelerator and experiments been successfully commissioned and operated?  {\bf Yes.}  

(2)  Have we created a state of matter that is not hadronic?  {\bf Yes.}  

(3)  Have we created a weakly interacting gas of quarks
and gluons (``the Quark Gluon Plasma'')?  {\bf No.  }

(4)  Have we created a strongly interacting partonic system (a different type
of quark gluon plasma)?  {\bf Yes.}  However, in answering yes to the last question, we are only at the start of understanding
its properties at a truly quantitative level.  

This proceedings is by no means a complete summary of the state of the field.  Additionally, there is an emerging area of interest
involving the physics of gluon saturation.  A good summary of this physics and recent results are found in.\cite{mclerran}

\section{Expectations}

\begin{figure}
\centerline{\psfig{file=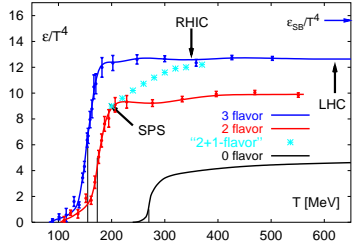,width=5cm}}
\vspace*{8pt}
\caption{Lattice QCD results for the energy density / $T^4$ as a function of
the temperature ($MeV$).  Note the arrow on the right side indicating the
level for the Stefan-Boltzmann limiting case.\label{fig_lattice}}
\end{figure}

Lattice QCD predicts a transition to a quark gluon plasma at high temperature where the number of degrees of freedom is
significantly increased.  If we consider a non-interacting system of $g$ massless degrees of freedom, we find the following
relation:
\begin{equation}
\epsilon = g {{\pi^2} \over 30} T^{4}
\end{equation}
where $\epsilon$ is the energy density  and $T$ is the thermal temperature.  As shown in Figure~\ref{fig_lattice}, 
there is a rapid increase
in the number of degrees of freedom at the transitions value $T=170~MeV$ and energy density of order 1 $GeV/fm^3$.\cite{karsch}  
The energy density value at high temperature approaches
80\% of the non-interacting gas limit.  This observation has led many to conclude that we expect a ``weakly'' interacting
gas of quarks and gluons where the long range confining potential is screened in medium.

QCD in vacuum is characterized by a linear increase in the potential as a function of the distance between color charges.  This leads to the confinement
of quarks and gluons inside of hadrons, either as baryons or mesons.\cite{greiner}  There is spontaneous breaking of approximate chiral symmetry and the quarks
thus take on a mass much larger than the neutral-current masses.  QCD in very dense or very hot conditions is characterized by a screening of the color
charges and the potential vanishes for large distance scales.  One thus has a deconfinement of quarks and gluons from hadrons and a restoration
of approximate chiral symmetry.  

As noted before, the lattice results led many to believe the transition was to a weakly interacting gas of quarks and gluons.  In fact, many
calculations have been done in the weakly coupled or even non-interacting limit for simplicity.  We know that only at very high temperatures (greater
than 100 times the transition temperature) does one achieve asymptotic freedom of the partons.\cite{dokshitzer}  
However, even at these high temperatures, interactions
always have a large component at low $Q^2$ where the coupling constant $\alpha_S$ is still large.  In fact, at the temperatures we might achieve
at RHIC or the Large Hadron Collider (LHC), the strong coupling constant in vacuum is not small.  Thus, it is possible that we should expect
a strongly-coupled plasma.

\begin{figure}
\centerline{\psfig{file=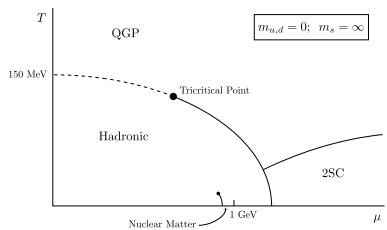,width=5cm}}
\vspace*{8pt}
\caption{Theoretical phase diagram of nuclear matter, not yet confirmed by experiment.  Note that this diagram is for the case of
two light quarks.\label{fig_phases}}
\end{figure}

The field of heavy ion physics is often described via a single plot of the phase diagram of nuclear matter as shown in 
Figure~\ref{fig_phases}.\cite{krishna}  In early versions, even
as recently as five years ago, the diagram had a phase boundary extending from the high temperature, low net baryon density domain to the high density,
low temperature domain where everything above that boundary was labeled as a quark gluon plasma.  In the last few years, major theoretical advances have
been made for the high density domain by Wilczek and Rajogopal.  Quark-quark interactions allow for an analog of Cooper pairs to form, creating 
non color-neutral quasi-particles.  They have thus renamed what was once quark gluon plasma to a Color Superconductor in the high density domain.
There may be a true phase boundary between this Color Superconductor and other regions of the phase diagram at higher temperatures.  In the high temperature
region our understanding is currently being driven by experimental results at RHIC.  For now, we will use the quark gluon plasma definition as
simply a new state of matter where the fundamental degrees of freedom are not color neutral hadrons.  Perhaps later we will come up with a more
exciting or descriptive name.

Quark-gluon plasma physics is relevant for the early universe a few microseconds after the Big Bang.  At that time all matter in the universe was in
this state and transitioned into confined partons in hadrons shortly thereafter.  Ed Witten wrote that: ``A first-order QCD phase transition that
occurred in the early universe would lead to a surprisingly rich cosmological scenario.''\cite{witten}  However, he did note that it is possible that signatures
from this transition may not have survived until today.  Perhaps a very inhomogeneous early universe due to bubble formation during the transition 
could affect the implications of Big Bang Nucleosynthesis (BBN).  Recent measurements from Boomerang~\cite{boomerang} and now WMAP indicate a very homogeneous universe
at the time of photon decoupling and they confirm a homogeneous universe hypothesis used to calculate light nuclei yields from earlier BBN.  It is possible
that there was no strong first-order transition or that diffusion erased any such signature.  Either way, it is clear that while quark gluon
plasma physics is interesting in understanding a particular epoch of the early universe, it is not a key part of understanding its future evolution.  Thus,
we must study this aspect of QCD with accelerators on Earth.

\section{Heavy Ion Collisions}

\begin{figure}
\centerline{\psfig{file=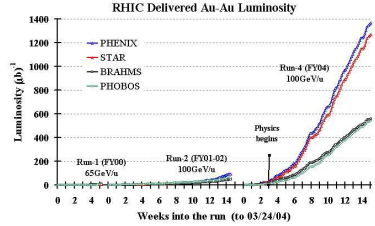,width=5cm}}
\vspace*{8pt}
\caption{RHIC delivered luminosity for gold-gold reactions as a function of time.\label{fig_rhic}}
\end{figure}

In a heavy ion collision at RHIC, of order 10,000 gluons, quarks and antiquarks from the nuclear wave-function are made physical in the 
laboratory!  The key question is what is the nature of this ensemble of partons?  

The Relativistic Heavy Ion Collider has been online since 2000.  The design Gold-Gold (Au-Au) energy and luminosity have been achieved as of 2004.  All 
experiments have successfully been taking data.  Additionally, the polarized proton-proton (spin) program is underway.  The luminosity achieved
in 2004 exceeded expectations, and the design value was not just reached but delivered over a large fraction of the running time.  The performance for heavy
ion beams is shown in Figure~\ref{fig_rhic}.

\section{Initial Conditions}

\begin{figure}
\centerline{\psfig{file=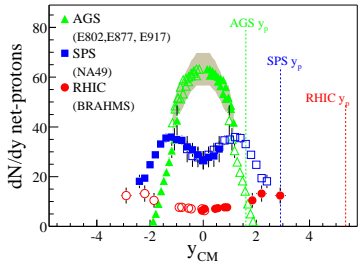,width=5cm}}
\vspace*{8pt}
\caption{BRAHMS experimental result for dN/dy of net-protons (protons minus antiprotons).  Also shown are results from lower energy experiments.\label{fig_stopping}}
\end{figure}

As an experimentalist, one should try to understand what one can learn from the data alone and then bring in more detailed model descriptions.  
At full RHIC energy, a total maximum energy of 39.4 TeV is colliding in central Au-Au reactions.  The BRAHMS experiment has measured the distribution of
net-protons (protons minus antiprotons), as shown in Figure~\ref{fig_stopping}.\cite{brahms_stopping}  
From this distribution one calculates that 26 TeV of energy is made available in the collisions for 
heating the vacuum and eventually producing new particles.   The remaining energy is maintained by fragments traveling along the beam direction.

From this 26 TeV of available energy, we want to know what energy density is achieved at the early time just after the collision.  The PHENIX
experiment has measured the resulting transverse energy~\cite{phenix_et} and using the Bjorken energy density equation~\cite{bjorken}:
\begin{equation}
\epsilon_{Bj} = {{1} \over {\pi R^2}} {{1} \over {2 c \tau}} \left( 2 {{dE_T}\over{dy}} \right)
\end{equation}
one determines the energy density as 23.0 $GeV/fm^3$ at a time 0.2 $fm/c$ after the reaction.  We expect from longitudinal expansion that this
energy density should drop at least as fast as $1/t$.  However, for a few $fm/c$ the energy density is well above the transition level predicted
from lattice QCD.

\section{Probes of the Medium}

We can also understand the density of the medium by sending calibrated probes through the system and determining its opacity.  The ``calibrated probes'' 
we use are quarks and gluons at high transverse momentum from hard parton-parton scatterings from the nuclear target and projectile.  We refer to these probes
as calibrated since we can calculate their rate in the framework of perturbative QCD, factorization and universality.  Full energy reconstruction of
jets resulting from quark or gluon fragmentation is good proxy for the underlying scattered parton.  At present the RHIC experiments measure the yield
of high transverse momentum hadrons, and not fully reconstructed jets.  
Thus, we must fold experimentally parameterized fragmentation functions and parton distribution functions, along
with the pQCD calculable scattering cross sections.  Results from the PHENIX experiment for the yield of neutral pions from $p_{T} = 2-13$ GeV/c
in proton-proton reactions agrees quite well with pQCD calculations.\cite{phenix_pp}  Thus, we can extend the calculation of expected rates to nuclear collisions
by scaling the yields with the nuclear thickness as a function of collision impact parameter.  

\begin{figure}
\centerline{\psfig{file=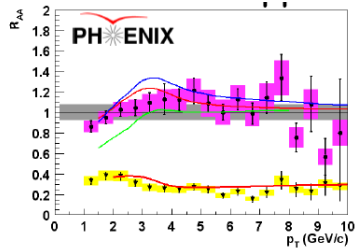,width=5cm}}
\vspace*{8pt}
\caption{PHENIX data on $R_{AA}$, the ratio of neutral pion yields in $AA$ reactions
relative to $pp$ reactions scaled by the nuclear thickness, shown as the lower triangles.  
The upper data points are $R_{dA}$ comparing yields in deuteron-gold reactions to proton-proton.\label{fig_raa}}
\end{figure}

\begin{figure}
\centerline{\psfig{file=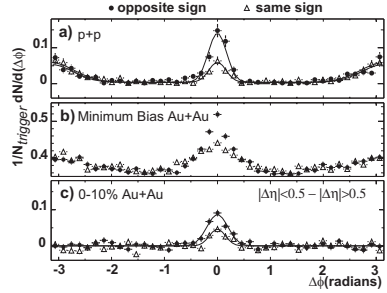,width=5cm}}
\vspace*{8pt}
\caption{Azimuthal distributions of same-sign and opposite sign pairs for a) p+p
b) minimum bias Au+Au and c) background subtracted central Au+Au collisions.  All
correlation functions require a trigger particle with $4<p_{T}^{trig}<6~GeV/c$ and 
associated particles with $2<p_{T}<p_{T}^{trig}~GeV$. 
\label{fig_star1}}
\end{figure}

What do we then observe in heavy ion reactions?  All four experiments now have results that indicate a large suppression (up to a factor of five) of
high $p_T$ hadrons relative to the scaled pQCD expectations, as shown from the PHENIX
experiment in Figure~\ref{fig_raa}.\cite{phenix_raa}  
In peripheral reactions (large impact parameter), there is no suppression, and the suppression
grows as one gets to more and more central (small impact parameter) reactions.  The STAR experiment has published the azimuthal angular distribution
of hadrons relative to the highest $p_T$ hadron (trigger particle) in the event.  In proton-proton reactions, one sees a near-side correlation 
(close in angle to the trigger particle) of other hadrons, from the fragmentation of the same parton.  One also observes an away-side correlation 
(180 degrees opposite to the trigger particle) from the partner scattered parton (which must be opposite in azimuthal angle by conservation of momentum,
assuming no initial $k_T$ or additional radiation).  Thus, these results confirm that the high $p_T$ hadrons are measuring the results of our
``calibrated probes'', partons propagating through the medium and then fragmenting.  In heavy ion reactions, for the most central gold-gold case, the
away-side jet correlation disappears for hadrons with $p_T>2$ GeV/c as seen in Figure~\ref{fig_star1}.\cite{starjet1}
 Conservation of energy and momentum says that the away side jet cannot really
disappear.  So the natural question is what has happened to this energy?  

In order to understand the answer to this question, we put forth a much simplified picture of the dynamics.  If high $p_T$ hadrons are suppressed or
absorbed in the medium, we should expect that they are more affected if their path length through the medium is longer.  Therefore, having a high $p_T$
trigger particle biases that particle for being emitted near the collision volume surface, and thus having a very short path through the medium.  This
short path bias has the opposite effect on the partner scattered parton, which is now biased for having a very long path through the medium.  In this 
picture, it is the trigger bias that results in almost no modification to the near-side correlation and a large modification to the away-side.  What
if the medium is very opaque to colored partons?  Perhaps the partons lose energy in medium and thus all their fragmentation products end up below
$p_{T}=2$ GeV/c.  Perhaps scatting of the partons or resulting hadrons causes a broadening of the angular distribution of hadrons.  Perhaps we have created
a black hole in these reactions that eats the partons?  The last one is just mentioned since this kind of speculation always seems to make the
local newspapers.  

\begin{figure}
\centerline{\psfig{file=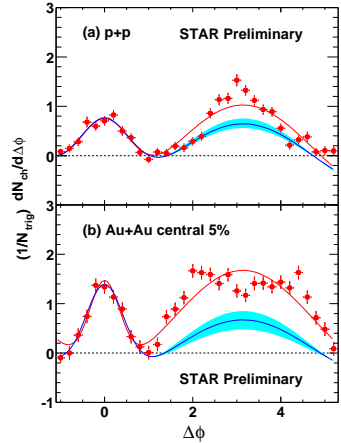,width=5cm}}
\vspace*{8pt}
\caption{Preliminary results from the STAR experiment on azimuthal distributions of
hadrons in proton-proton and gold-gold reactions.  The threshold for inclusion
of hadrons in the correlation distribution is $p_{T}>200~MeV$.
\label{fig_star2}}
\end{figure}

In this last year, new results from the STAR and PHENIX collaborations have answered these questions.  The PHENIX experiment has seen that with a lower
$p_T$ threshold on the correlation plot, one can see the away side jet but with a much broader angular distribution.\cite{phenixjet}  
The STAR experiment has lowered
their $p_T$ threshold down to 200 MeV/c and now sees an away-side correlation that is very broad and does not really resemble a jet-cone, shown in Figure~\ref{fig_star2}.\cite{starjet2}  
One recovers the ``lost'' energy and momentum in a broad distribution of hadrons with $<p_{T}> \approx 500$ MeV/c.  This momentum is not very different
from the thermal temperature of the medium.  One can thus think of a 10 GeV parton losing energy in medium and having it approximately thermalized!

\begin{figure}[ht]
\centerline{\epsfxsize=4.1in\epsfbox{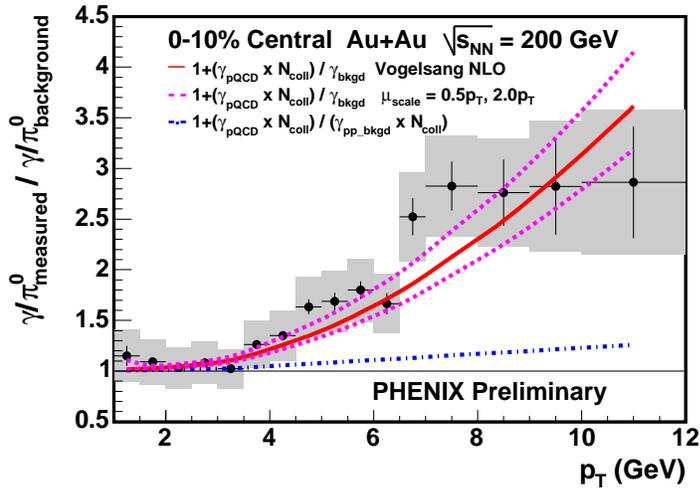}}
\caption{PHENIX preliminary direct photon results in central gold-gold reactions. 
The data is plotted as a ratio of photons to neutral pions for data compared to
simulation (including no direct photons).   The observed excess is consistent
with NLO pQCD expectations for direct photons scaled by the nuclear thickness, but
with neutral pions suppressed as previously measured by PHENIX.\label{fig_photon}}
\end{figure}

It would be very useful to have another probe of the medium that is also sensitive to the initial flux of incident partons (nuclear parton
distribution functions) and to pQCD calculable scattering rates, but that does not ``see'' this opaque medium.  Direct photons from gluon-Compton
scattering are a perfect fit for what we want.  In particular, after the quark-gluon scattering, the photon essentially does not ``see'' the medium
because it does not couple to any of the color charges.  Its mean free path from electromagnetic interactions is many times longer than the nuclear
size.  PHENIX has results on direct photon production, as shown in Figure~\ref{fig_photon}\cite{justin}, and these photons appear to follow the pQCD expected rates (within 30\%) with no medium modification, as expected.  

This experimentally observed phenomena was in fact predicted and is given the name ``jet quenching.''  Partons are expected to lose energy via
induced gluon radiation from multiple scattering as they traverse the dense color charge medium.  Also, coherence among these radiated gluons
may lead to an energy loss proportional not to the path length, but to the path length squared.  This results in an effective softening of the
parton fragmentation function.  It is almost impossible to have most of the original parton energy go to one hadron, since the parton has already
lost substantial energy via additional gluon emission before the final fragmentation.  There are a few variants of such calculations which differ 
in their details.  We show in Figure~\ref{fig_raa} a comparison with the GLV formalism\cite{glv} which agrees quite well with the data assuming an initial 
energy density of color charges of order $15~GeV/fm^3$.  Note that these calculations are sensitive to the lowest energy radiation and thus the
infrared cutoff scale.  In the GLV formalism, they connect this cutoff scale to the natural plasma frequency.  

An additional test of this formalism will be the measurement of high $p_T$ charm and beauty hadrons.  QCD is flavor independent, but heavy quarks
at the same $p_T$ as light quarks are moving at a much slower velocity.  It is predicted that there should be a dead-cone without radiation
in the forward direction relative to the parton, thus ensuring that emitted radiation does not violate causality.\cite{dima,magdelena}  Recent single lepton from 
semi-leptonic decays of D and B mesons from STAR and PHENIX, and complete D meson reconstruction from the STAR experiment, should be able to address this question.
\cite{kelly,starcharm}
Recently, we have put forth a proposal for measuring three-jet events in heavy ion reactions at the Large Hadron Collider with the ATLAS detector.\cite{nagle_3jet}
These topologies are guaranteed to have gluon jets, and thus one might statistically isolate gluon jets.  The gluon having two colors (e.g. red-antigreen) 
should have twice as strong a coupling to the medium and lose substantially more energy than quarks.

\section{Baryon Issue}

\begin{figure}
\centerline{\psfig{file=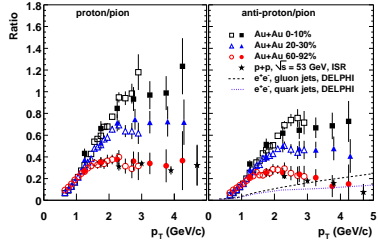,width=5cm}}
\vspace*{8pt}
\caption{PHENIX ratio of (anti) protons to pions as a function of $p_{T}$ for
various centralities.\label{fig_ratios}}
\end{figure}

There is one result that at first appearance seems to disagree with the previous picture.  If the parton radiates gluons first and then has 
a final fragmentation in vacuum, the ratio of species of high z hadrons should look like vacuum fragmentation.  However, the RHIC experiments have
observed a large enhancement of baryons and antibaryons relative to mesons at intermediate $p_T=2-6$ GeV/c.  In proton-proton reactions, the ratio of 
antiprotons to negative pions is of order 1/3.  However, in central gold-gold reactions, the ratio reached a value of order one as shown in 
Figure~\ref{fig_ratios}.\cite{phenixratio}
Similar results are observed for lambdas and kaons with the ratio returning to vacuum fragmentation expectations only above 5-6 GeV/c.  It has been proposed
that in this kinematic range that hadron production is not dominantly from fragmentation, but rather recombination of partons from different
sources (some of them being thermal).  These color recombination models are currently being tested through detailed comparison with experimental
data (for example\cite{recomb}).

\section{Collective Motion}

Thus far we have determined that the system created is very opaque and has a high energy density.  But we have only briefly
commented on the issue of thermalization (i.e. collectivity).  

In non-central nucleus-nucleus reactions, there is a large spatial anisotropy of the collision volume (i.e. it is ellipsoidally shaped).  
The degree to which this spatial anisotropy is translated into momentum space is an excellent measure of the thermalization and pressure.
All experiments observe a large momentum space anisotropy which can be characterized via a Fourier decomposition of the azimuthal angular
distribution of hadrons.  The strongest component is termed $v_2$ and corresponds to differences in particle emission along the minor
axis of the ellipse relative to the major axis of the ellipse.  

\begin{figure}
\centerline{\psfig{file=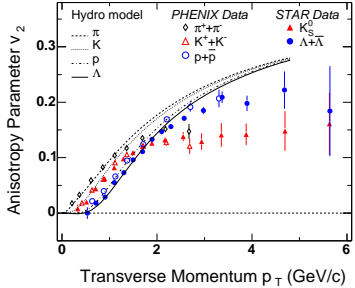,width=5cm}}
\vspace*{8pt}
\caption{Anisotropy parameter $v_{2}$ as a function of $p_{T}$ for various
hadron species.  Also shown are results from a hydrodynamic model calculation.\label{fig_flow}}
\end{figure}

As shown in Figure~\ref{fig_flow}, all hadrons including the multiply strange baryons have a large flow $v_2$ component, indicating very large
re-scattering.\cite{starflow,phenixflow}  
A preliminary result from the PHENIX experiment may even indicate collective motion of charm particles.\cite{kelly,batsouli}  
Understanding the experimental data often utlitizes hydrodynamic model calculations.  These calculations assume early equilibration and start with an 
initial geometry (energy density and spatial configuration) determined from the nuclear geometry and impact parameter).  They then follow 
a set of equations of motion and utilize an equation of state derived to match that from lattice QCD results.  Typically calculation are
done assuming complete equilibration and zero viscosity (resistance to shear forces and hence to flow).  As shown in Figure~\ref{fig_flow}, there is
reasonable agreement between the experimental data for the low transverse momentum distribution of hadrons and their $v_2$ near
mid-rapidity and the calculations.  This is the first time hydrodynamics with zero viscosity has been able to describe heavy ion 
reactions to this degree of accuracy.  Note that these calculations typically imply initial conditions with a thermalization
time of order 0.2-2.0 fm/c and an energy density at that time of 10-20 $GeV/fm^3$ (for example\cite{kolb}).   

This is very strong evidence for a large degree of thermalization.  Only a very strongly coupled system can approach
equilibrium this quickly.  It should be noted however, that details of the hydrodynamic description still need to be worked out.  
They currently do not describe the longitudinal distribution of $v_2$.\cite{phobos_flow}.
Thus, an conclusions about exact latent heat or limits on the viscosity should be viewed with some skepticism.


\section{Partonic or Hadronic?}

High energy proton-proton reactions cannot be described in terms of purely hadronic degrees of freedom.  
Phenomenological descriptions have included color string models for example.  These reactions do not
create a quark gluon plasma because strings fragment into hadrons without time for thermalization
(i.e. re-interaction).  Transport models (such as UrQMD, HSD, RQMD, and others) have been developed for
nuclear collisions that include color strings, resonances and hadronic interactions.  This class of models
under predict the collective motion measured by $v_2$ by a factor of 4-10.\cite{stocker}  Only if we violate quantum 
mechanics and allow the hadronic wave-functions to fully form in zero time can they come close to the data. 
Thus, there must be substantial interaction amongst non-hadronic particles.  

\section{Quark Gluon Plasma}

There are many indications of energy density values well above the transition level predicted by
lattice QCD. However, the system does not behave anything like a ``weakly'' interacting gas of
quarks and gluons.  It is more like a liquid (zero mean free path) with very low viscosity (zero
resistance to flow).  What are the new quasi-particles or parton level couplings that lead to such
strong collectivity?  We have seen in recent lattice QCD calculations that some bound states, such as
charmonium, may persist up to 1.5 $T_c$, and vector meson states have modified spectral functions but
remain as correlations between partons.  Further studies are needed and most of all experimental identification of these
states, possible through the measurement of low mass dileptons at RHIC.


\section{Phase Transition}

Many early calculations related to heavy ion reactions predicted discontinuities in various observables as a function of collision conditions.  
These predictions were based on the relatively sharp transition as a function of temperature seen in the lattice QCD results.  These were
the advertised ``smoking gun'' signatures of the quark gluon plasma transition.  

A single heavy ion collision does not measure at one temperature or even one energy density.  Rather it measures an energy
density profile including a significantly more diffuse corona.  The system expands too quickly for nucleation from the center region.
Thus, it is unlikely to see a true discontinuity in any observable as a function of collision energy or collision centrality (impact
parameter).  What does this mean for the field?  It means that we have had a much more challenging task to demonstrate the
properties of the created system, rather than just one ``smoking gun'' discontinuity observable.  The RHIC community is meeting
this challenge. 

\section{Conclusions}
The RHIC program is operating very successfully.  The gluon and energy density of the created matter is well above the
lattice QCD predicted transition level.  The matter is behaving collectively as a very small viscosity liquid.  This is not
the traditionally thought of weakly interacting gas of quarks and gluons (``the Quark Gluon Plasma'').  However, this is
the creation of a strongly interacting quark gluon plasma (or the ``Quark Gluon Liquid'').  Direct experimental identification
of the new quasi-particles and medium properties (such as the exact viscosity) remains a challenge for the near term
future.  The results from the last few years of running have been very exciting and we expect that to continue over the next
decade.

\section*{Acknowledgments}
We are very grateful to the conference organizers for the opportunity to discuss heavy ion physics in the context of
this high energy particle physics meeting.  We acknowledge support from the Department of Energy (DE-FG02-03ER41244) and the
Alfred P. Sloan Foundation.

\end{document}